# *T*-linear specific heat in pressurized and magnetized Shastry-Sutherland Mott insulator SrCu$_2$(BO$_3$)$_2$

Jing Guo[1,2†], Pengyu Wang[1,2†], Cheng Huang[3,4†], Chengkang Zhou[3,4†], Menghan Song[3,4†], Xintian Chen[1,2], Ting-Tung Wang[3,4], Wenshan Hong[1], Shu Cai[1,5], Jinyu Zhao[1,2], Jinyu Han[1,2], Yazhou Zhou[1], Qi Wu[1], Shiliang Li[1,2], Zi Yang Meng[3,4*], Liling Sun[1,2,5*]

[1]Beijing National Laboratory for Condensed Matter Physics and Institute of Physics, Chinese Academy of Sciences, Beijing 100190, China

[2]University of Chinese Academy of Sciences, Beijing 100049, China

[3]Department of Physics and HK Institute of Quantum Science & Technology, The University of Hong Kong, Pokfulam Road, Hong Kong SAR, China

[4]State Key Laboratory of Optical Quantum Materials,

The University of Hong Kong, Pokfulam Road, Hong Kong SAR, China

[5]Center for High Pressure Science & Technology Advanced Research, Beijing 100094, China

†These authors contributed equally to this work.
*Corresponding authors: Email: zymeng@hku.hk, llsun@iphy.ac.cn or liling.sun@hpstar.ac.cn

**The pressurized Shastry-Sutherland Mott insulator SrCu$_2$(BO$_3$)$_2$ has been found to host a plaquette-singlet phase and an antiferromagnetic phase that break different symmetries spontaneously** [Nat. Phys. **13**, 962 (2017)，Phys. Rev. Lett. 124, 206602 (2020)]**. The recent experiment showed that their transition is of a first order nature** [Commun Phys 8, 75 (2025)]**, which seems against the pursuit of exotic and deconfined degrees of freedom in this famous frustrated quantum magnet. We found a new direction in this study. By applying a magnetic field to the material, we discover that SrCu$_2$(BO$_3$)$_2$ exhibits a universal and metallic *T*-linear specific heat behavior in a large magnetitic field range close to the pressure of zero-field first order transition between plaquette-singlet and antiferromagnetic phases. Such an unexpected gapless response from an electronically gapped Mott insulator could be attributed to magnetized Dirac spinons liberated by the combined effect of magnetic field and pressure, consistently seen from our quantum many-body thermal tensor network computation of the Shastry-Sutherland model under**

**magnetic field. Such a robust and universal *T*-linear specific heat phase points out the richness of the phase diagram of the material expanded by the axes of pressure and magnetic field and is calling for new theoretical frameworks to its full explanation.**

## Introduction

In the past two decades, great efforts have been devoted to detecting the deconfined quantum critical points (DQCPs) and the associated quantum spin liquids (QSLs) with fractional spinon excitations that are beyond the conventional Landau-Ginzburg-Wilson paradigm[1–3] at the theoretical and computational levels of quantum spin models[4-25]. However, their experimental realizations and detections in magnetic Mott insulators are rare. Among a few possible candidates[26-34], extensive investigations have been performed on a magnetic Mott insulator $SrCu_2(BO_3)_2$ to detect whether there is a DQCP or a QSL[27-31] state, as the frustrated spin interactions of this Mott insulator resembles largely those of the famous Shastry-Sutherland (SS) model (with $J'$ and $J$ referring to the intra- and inter-dimer Heisenberg spin couplings)[35], which hosts a dimer-singlet (DS) state, a plaquette valence-bond state or plaquette singlet (PS), an antiferromagnetic state (AF) at different $J/J'$[36–44], and more recently, a putative QSL in a very narrow window of $J/J'$ between PS and AF[45-47].

The PS and AF phases spontaneously break different symmetries: PS breaks lattice translation symmetry and AF breaks spin rotational symmetry. Fractional excitations and the associated emergent gauge field[3,5,6,48] are needed to describe such a PS-AF transition, if, by tuning pressure, a DQCP or even a QSL phase is to emerge [30,38-40,45-47]. However, the latest pressurized specific heat measurement showed that the zero-field PS-AF transition is of first-order nature[31]. Nevertheless, strong quantum fluctuations around the first-order transition point signified by the residue specific heat at extremely low temperature are observed, so that the DQCP or a QSL could be blockaded by a first-order energy barrier. Moreover, previous nuclear magnetic resonance[30] studies on

SrCu$_2$(BO$_3$)$_2$ with out-of-plane magnetic fields at the PS pressure range also suggested that the PS-AF transition with enhanced fluctuations. In addition, magnetic plateaus are found at high magnetic fields in the ambient-pressure DS phase of SrCu$_2$(BO$_3$)$_2$[49]. All the evidence suggests the investigation under magnetic field of the pressurized material close to the zero-field PS-AF transition, which has not been performed so far, might lead to new phases and transitions, and these are the major discoveries of this work.

In this study, we performed specific-heat measurements under simultaneous high pressure and magnetic field to map the thermodynamic response across the vicinity of PS-AF transition (see Methods and Supplemental Material (SM)[50] for details). We uncover a robust and universal *T*-linear specific-heat region emerging at low temperatures with the maximum magnetic field span close to the pressure of zero-field PS-AF transition. Optical observations confirm that the material remains insulating behavior throughout the pressure range investigated, indicating frozen charge degrees of freedom and pointing out the exotic nature of the gapless excitations in this Mott insulator. We also noted the possibility of other extrinsic effects that causes gapless behavior, such as disorder and glassy behavior in our hydrostatic-pressure measurements, are very low, however we observed the universal gapless behavior in a wide range of pressure and magnetic field (see more below).

The *T*-linear specific heat of such a Mott insulator with no obvious order thus arises from gapless magnetic degrees of freedom, such as the emergent Dirac spinons which are fractionalized excitations, and our experimental observation likely suggests the presence of Dirac spinon state (DSS) of the material under the magnetic field with the fractionalized spinons forming Fermi surface. Moreover, we also performed the state-of-the-art quantum many-body thermodynamic calculations with the exponential thermal tensor network renormalization group (XTRG)[58] on the spin-½ Shastry-Sutherland model under magnetic field and found consistent gapless behavior in specific heat in the parameter regime in the vicinity of the PS-AF transition under finite Zeemann field[59]. Our experimental discovery and model calculation therefore open the door to investigations of the possible liberation of deconfined magnetized Dirac spinons by the competing interactions due to the combined effect of magnetic field and pressure in this famous Shastry-Sutherland Mott insulator $SrCu_2(BO_3)_2$.

## Results

The specific-heat measurements under different pressures and magnetic fields are presented in Fig. 1 and Fig. S5 in SM[50]. At 2.3 GPa and fields below 5.5 T (Fig. 1 a), the critical temperature of PS phase ($T_{PS}$) shifts to lower temperatures as the magnetic field increases, consistent with previous results[29,30]. The intensity of the PS peak diminishes with an increasing field and becomes undetectable when the magnetic field reaches 5.5 T. Afterwards, a new peak that indicates a magnetic-field-induced antiferromagnetic phase[29,30], AF′, is forming. Its transition temperature $T_{AF'}$ moves to higher temperatures as the field increases (as shown in Fig. 1 (b)). The suppression of the PS phase and the formation of the AF′ phase is consistent with NMR studies at comparable pressure[30], indicating that the thermodynamic features we observed originate from the same underlying microscopic spin reorganization. At 2.56 GPa, the PS-AF′ exhibits a pronounced first-order transition at ~1.3 K around 4.2 T (as shown in Fig. 1 (c) and (d) and Fig. S5 (b)). The PS anomaly persists just below 4 T but is abruptly replaced by a strong AF′ peak at 4.5 T, signaling the collapse of the PS phase and the onset of AF′. Compared with the results observed at 2.3 GPa, the $T_{AF'}$ peaks are sharper, indicating a more distinct transition under the increased pressure.

From 2.68 GPa onwards, there are no reported measurements under magnetic field. We found that the PS state is suppressed at a very low magnetic field (~1 T, see Fig. 1 (e)). Between 1 T and 4 T, no peak corresponding to the PS or AF′ state is detected down to ~ 0.4 K. In the magnetic field range of 1.5 - 4 T, the specific heat $C$ exhibits a $T$-linear behavior below 1.5 K with a cutoff of around 1.1 JK$^{-2}$mol$^{-1}$ for $C/T$ (Fig. 1 (f) and labeled by the dashed line and the yellow shadow), making a notable difference from the phase diagrams at 2.3 GPa and 2.56 GPa between the PS and AF′ phases. This $T$-linear behavior disappears above 4 T, where the AF′ peaks start to appear.

Increasing the pressure further to 2.75 GPa, corresponding to the pressure of zero-field first-order PS-AF transition[31], the $T$-linear behavior exists in an even broader magnetic range of 3 - 7 T, as labeled by dashed lines and yellow shadows in Fig. 1 (g) and (h), followed by the appearance of the AF′ phase starting at a higher field (9 T), indicating that the system changes significantly under pressure and magnetic field. The linear specific heats $C/T$ exhibit an intercept universally around 1.1~1.5 JK$^{-2}$mol$^{-1}$. Meanwhile, the material is insulated throughout, as confirmed by the optical observations (Sec. 5 and Fig. S6 of the SM[50]), where the sample consistently retains its translucent blue color, identical to that at ambient pressure. This indicates that the charge degrees of freedom in the material are frozen, and the system remains a Mott insulator. We note that the possibility of disorder and glassy behavior that could give rise to the observed gapless behavior is extremely low. As in those cases, the $C/T$ could give rise to very different low temperature behavior (for example, different intercept) at different magnetic fields and pressures, but we didn't observe them. Therefore, the observed apparently gapless $T$-linear specific heat cannot originate from charge excitations, or very unlikely due to impurities or other extrinsic effect, but instead arise from spin or magnetic degrees of freedom in such a Mott insulator.

At the pressure of 2.9 GPa, according to the previous zero-field results[31], the zero-field ground state has changed to the pressure-induced AF phase. As shown in Fig. 1 (i), the phase transition temperature of the AF phase decreases with increasing magnetic fields below 2 T. At this pressure, the $T$-linear range of the specific heat shrink significantly as we can observe it within a small range of 2.5 - 4 T, still with intercept universally around 1.1 JK$^{-2}$mol$^{-1}$ as also indicated by the dashed line in Fig. 1 (i) and (j). This may be due to the influence of the nearby probable DSS at 2.75 GPa getting weaker. When the magnetic field is larger than 4.25 T, a clear low-temperature AF′ peak appears (see Fig. 1 (j)).

At 3.2 GPa, no $T$-linear behavior is observed at any fields, and the transition changes to the AF-AF′ type (Fig. 1 (k) - (l)). The transition temperature of the peaks related to AF decreases slightly with increasing magnetic field below 2.0 T, while the peak intensity weakens and broadens. Above 2.0 T, the AF′ phase emerges and its peak shifts to higher temperatures with elevating magnetic field, accompanied by enhanced peak intensity (see Fig. 1 (l)). We note that all the linear specific heats $C/T$ exhibit an intercept around 1.1~1.5 JK$^{-2}$mol$^{-1}$ universally, as shown in the panels of Fig. 1 (e), (f), (g), (h), (i) and (j). This universal gapless behavior is extremely difficult to understand from extrinsic effects such as disorder and glassy behavior.

To better visualize the data, we prepare the color mapping of $C/T$ versus temperature and magnetic field, shown in Fig. 2, with transition points labeled, to find the boundaries of phases. At 2.3 GPa (Fig. 2 (a)), ground-state phases are changed from PS to AF′ by the magnetic field, and a very low temperature transition is implied at 5.2 T, in agreement with previously reported NMR experimental results[30]. However, as the pressure increases to 2.56 GPa (Fig. 2 (b)), a sudden disappearance of the PS state, via first order transition as mentioned above, is replaced by the emergence of the AF′ state at nearly the same temperature, around 1.2 K, similar with the zero field first-order PS-AF transition[31]. At 2.68 GPa (Fig. 2 (c)), the gapless intermediate DSS is observed between 1.5 T and 4 T, with the $T$-linear behavior most clearly emerging at 3 T. We use a green colored text to indicate this DSS. At 2.75 GPa (Fig. 2 (d)), the DSS extends over a broader magnetic field range (3 T to 7 T), indicating that higher pressure stabilizes this phase. This maximum range corresponds to the point where the zero-field PS-AF transition occurs, suggesting a close relationship between the $T$-linear DSS and the PS-AF transition. As pressure increases beyond 2.75 GPa, the $T$-linear state range gradually shrinks. At 2.9 GPa (Fig. 2 (e)), the $T$-linear state is confined to a field range of 2.5 T to 4 T. This middle $T$-linear state completely vanishes at 3.2 GPa (Fig. 2 (f)). The mapping of these transitions across multiple pressures and magnetic fields allows for a comprehensive understanding of the phase diagram of $SrCu_2(BO_3)_2$, shedding light on the complex interplay among pressure, magnetic field and temperature, and thus providing new insights into the nature of the intermediate DSS state.

The connections of these color mappings to the zero-field phase diagram are shown in Fig. 3 (a) - (g). It is seen that the $T$-linear specific heat DSS arises by the magnetic field from the first-order PS-AF transition, showing there are some barriers to block the deconfined traits of the PS-AF transition, and magnetic field can help to remove these barriers. The low-temperature phase diagram versus magnetic field strength and pressure is shown in Fig. 3 (h), which demonstrates a vast area of the $T$-linear specific heat phase. We believe the metallic DSS in such a Mott insulator is originated from the deconfined spinons with emergent gauge field under magnetic field[3,5,6,48]. Such a state is clearly beyond the conventional phases of symmetry-breaking, its fully understanding on which posts a new question to the community.

**Discussion**

To further understand the experimental phase diagram, we performed the state-of-the-art quantum many-body thermodynamic simulation on the spin-½ antiferromagnetic Heisenberg model on the Shastry-Sutherland lattice. We note that the zero-field phase diagram have been computed with exact diagonalization, tensor-network and neural quantum state methods[20,37,39,45-47,60,61], with the DS, PS and AF phases successfully captured in the model calculation, but the thermodynamic phase diagram under magnetic field has only been obtained very recently by some of us[59]. Using the XTRG thermal tensor-network method[58], we find that the $T$-linear specific heat DSS is also observed in the Shastry-Sutherland lattice under a Zeeman field. To one's surprise, that the parameter range in which the DSS appears is also similar – in the vicinity of the pressure range of PS-AF transition, and let's explain the results below.

The Hamiltonian is given by:

$$H = J \sum_{\langle i,j \rangle} S_i S_j + J' \sum_{\langle\langle i,j \rangle\rangle} S_i S_j - h \sum_i S_i^z. \quad (1)$$

Here, $J$ refers to the inter-dimer coupling and $J'$ to the intra-dimer coupling, represented by the black and green bonds in Fig. 4 (a), respectively. We note $g = J/J'$ to mimic the effect of pressure and it is empirically known that the relation between $J$, $J'$ and the pressure are $J'(P) = [75 - 8.3P/\text{GPa}]$ K and $J(P) = [46.7 - 3.7P/\text{GPa}]$ K. The energy scale is accordingly set to $J' = 52.756\ K$ at 2.68 GPa[27,28]. Our calculations are performed on a $3 \times 12 \times 4$ lattice with periodic boundary conditions along the $y$-direction and open boundaries along the $x$-direction, as shown in Fig. 4 (a). The XTRG simulations are implemented using $U(1)$ symmetry with a bond dimension $D = 800$, maintaining a truncation error of approximately $10^{-2}$. The computation details are presented in the "Exponential Thermal Tensor Network (XTRG) calculation" section in the Methods. The calculations are carried out for three parameter sets: $g = 0.70, h = 1.6$ ($T$-linear specific heat DSS in Fig. 3), $g = 0.70, h = 0.0$ (PS phase in the zero-field case), and $g = 0.60$, $h = 0.0$ (DS phase in the zero-field). The computed specific heat divided by temperature, $C/T$, is plotted as a function of $T$ in Fig. 4 (b). For $g = 0.70$ and $h = 1.6$, the lowest three temperature points are used to perform a linear extrapolation to the zero-temperature limit based on the formula $C/T = a_0 + a_1 T$. Further technical details and additional data are presented in the Method.

In Fig. 4 (b), the $C/T$ curve at $g = 0.70$ with $h = 1.6$ increases as the temperature decreases, reaching a plateau below $T = 3$ K, a region highlighted with yellow shading. This temperature scale is roughly consistent with our experimental observations, which fall within the same order of magnitude: the shaded regions appear around 1.7 K in Fig. 1 (f) (2.68 GPa, 3 T), 1.5 K in Fig. 1 (h) (2.75 GPa, 4.0 T), and 1.0 K in Fig. 1 (i) (2.9 GPa, 3.5 T). Moreover, such a *T*-linear dependence is absent in the zero-field cases ($g = 0.60$ and 0.70, shown as blue circles and orange triangles, respectively). At g = 0.6 the model is inside the DS phase, and the computed *C/T* goes to zero quickly (exponentially) as reducing temperature. And at g = 0.7 the model is inside the PS phase; the computed *C/T* also goes to zero quickly with a smaller gap, and it is very consistent with the experimental observation at 2.3 GPa and 2.56 GPa in Fig.1 (a) and (c). At $g = 0.7$ but under magnetic field, the linear *C/T* is clearly seen; this behavior indicates the emergence of Dirac spinons activated by the applied field. The underlying mechanism of such DSS phase, with emergent spin degrees of freedom under magnetic field, as discussed above, post a new and exciting question to the community.

## Conclusion

Our experimental results and model computations find a robust and universal *T*-linear specific heat DSS state, which exists in a large area of the pressure and magnetic field, with the maximum *B* range locating at the pressure where the zero-field first-order PS-AF transition happens, as shown in Fig. 3 (h). Therefore, in such an insulating material, it is suggested that Dirac spinons as the fractional excitations emerge and proliferate in the parameter space of magnetic field and pressure by our experimental and computational results. The phase diagram illustrates the evolution of the phase boundaries as functions of both pressure and magnetic field, providing a comprehensive understanding of how the magnetic properties of the system are tuned by external conditions.

In a broader perspective, so far, the emergent Dirac spinon signature have been suggested to be observed in the inelastic neutron scattering (INS) in Kagome antiferromagnet YCu$_3$(OH)$_6$Br$_2$[Br$_{(1-x)}$(OH)$_x$][32-34], triangular antiferromagnets YbZn$_2$GaO$_5$[62,63] and $A$YbSe$_2$ ($A$ being an alkali metal)[64,65], and the putative quantum oscillation at the 1/9 magnetic plateau of the Kagome material[66,67]. SrCu$_2$(BO$_3$)$_2$, on the other hand, offers a new alternative to generate Dirac spinons. Our results offer very plausible evidence of the emergent Dirac spinons and the fractionalized quantum fluctuations in SrCu$_2$(BO$_3$)$_2$.

Complementary probes such as neutron scattering, magnetic torque, susceptibility, and NMR could further elucidate the nature of this spinon state and the associated deconfined quantum behavior. Theoretically, such an unexpected DSS phase seen from our state-of-the-art thermal tensor-network computation -- bears an astonishing similarity to the experimental findings in the material -- opens the door to further investigations of the possible liberation of deconfined magnetized Dirac spinons by the competing interactions in this highly frustrated quantum magnet model, and by the combined effects of magnetic field and pressure in the associated Shastry-Sutherland Mott insulator SrCu$_2$(BO$_3$)$_2$.

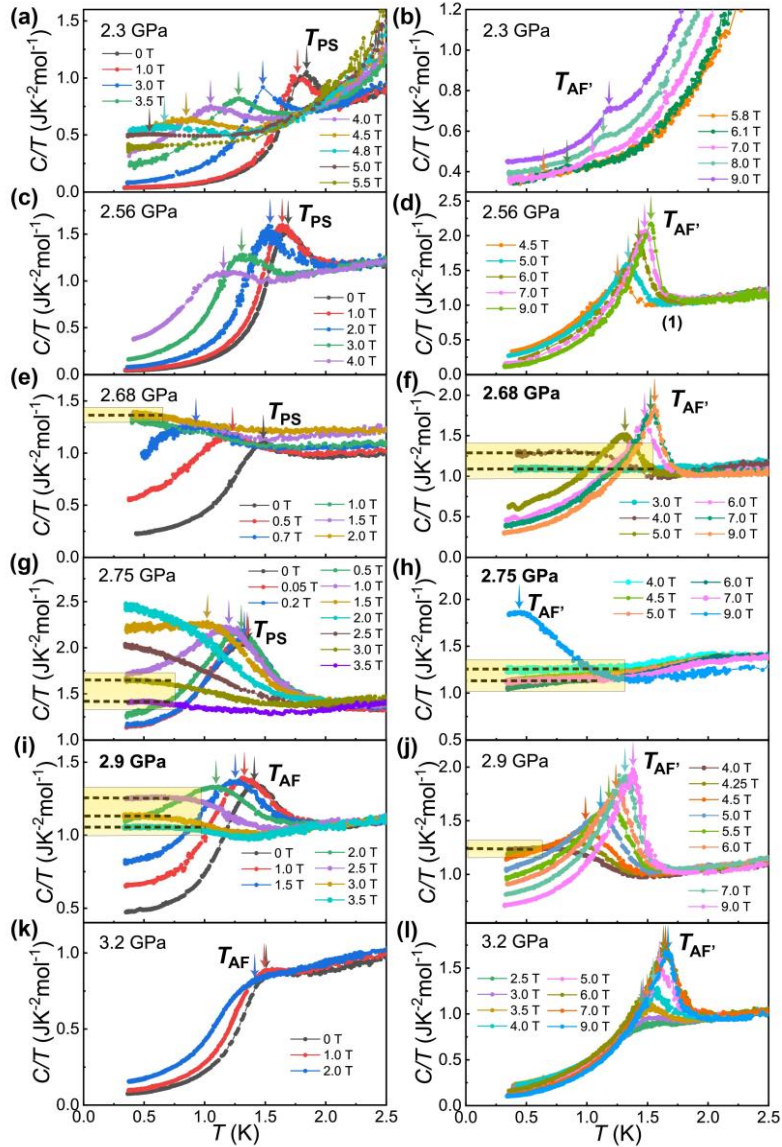

**Fig. 1. Evolutions of the specific heat capacity $C/T$ of SrCu$_2$(BO$_3$)$_2$ as a function of magnetic field and temperature for pressures ranging from 2.3 GPa to 3.2 GPa.** $C/T$ versus $T$ measured at 2.3 GPa (**a**), 2.56 GPa (**b**), 2.68 GPa (**c**), 2.75 GPa (**d**), 2.9 GPa (**e**) and 3.2 GPa (**f**). $T_{PS}$, $T_{AF'}$, and $T_{AF}$ stands for transition temperature of plaquette singlet (PS) state, magnetic-field-induced anti-ferromagnetic (AF′) state, and pressure-induced anti-ferromagnetic (AF) state respectively. $T$-linear specific heat regions are denoted by dashed lines and yellow shadows.

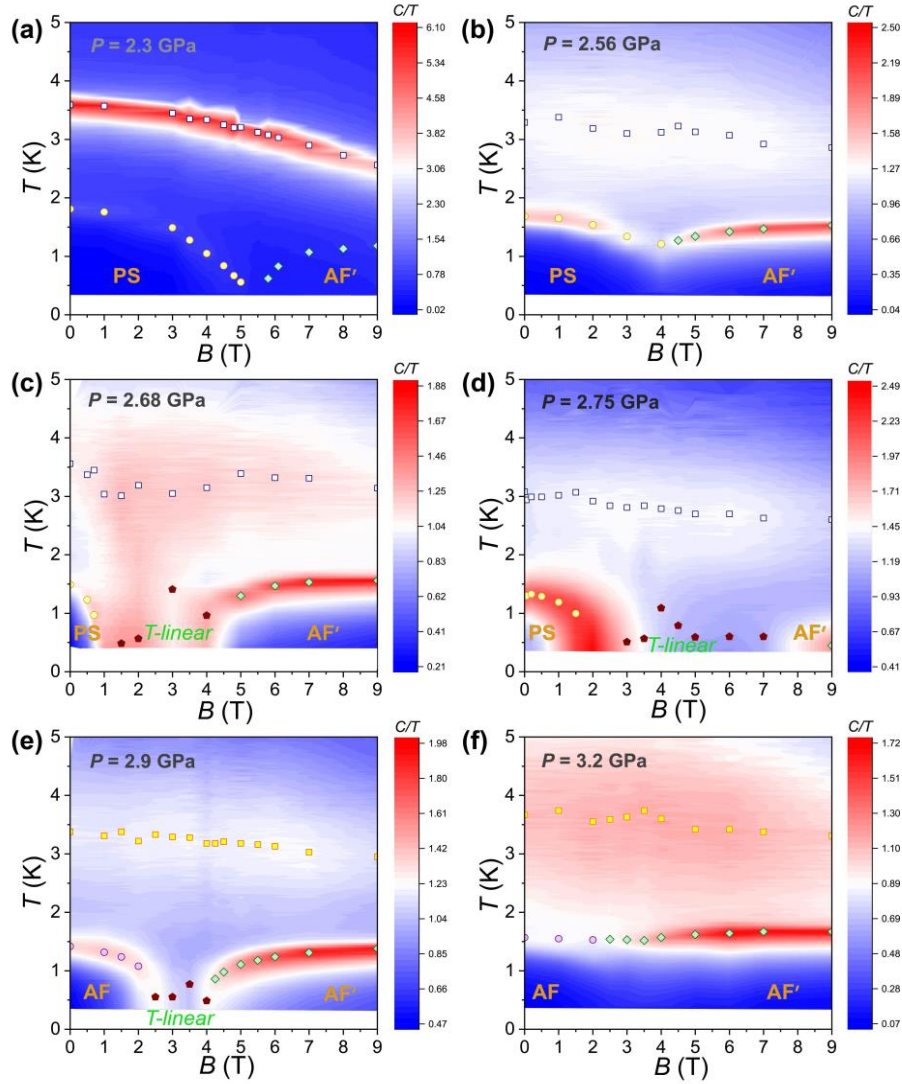

Fig. 2. **Mapping of the specific heat capacity over $T$ ($C/T$) of SrCu$_2$(BO$_3$)$_2$ as a function of temperature and magnetic field at various pressures.** (a) - (f) display the $C/T$ dependence of temperature and magnetic field at pressures of 2.3 GPa, 2.56 GPa, 2.68 GPa, 2.75 GPa, 2.9 GPa, and 3.2 GPa, respectively. The continuous color maps are generated by interpolating the measured $C/T$ curves to visualize the evolution of specific heat with temperature and magnetic field. The color scale represents the magnitude of $C/T$, while the green-texted regions are only to indicate the areas of $T$-linear behavior. The blue or yellow squares indicate the temperatures at which the correlations of the corresponding ground-state orders start to grow. Yellow circles represent the transition temperatures of the PS state. Green diamonds denote AF′ transition temperatures, and purple circles highlight low magnetic field AF transition temperatures.

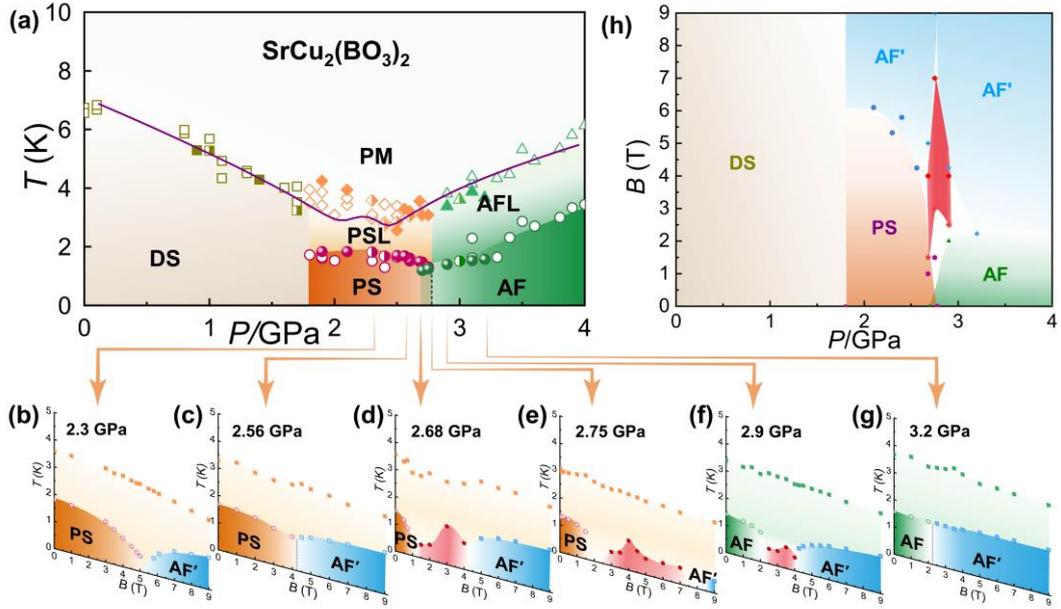

Fig. 3. **The pressure and magnetic field phase diagram of SrCu$_2$(BO$_3$)$_2$.** (a) The zero-field pressure-temperature (*P-T*) phase diagram. Here the acronyms AF′ and PM stand for the field-induced anti-ferromagnetism and para-ferromagnetism respectively. PSL and AFL stand for the correlation forming of the corresponding orders of PS and AF respectively, with L stands for liquid. Magnetic-field-temperature (*B-T*) phase diagrams at 2.3 GPa (b), 2.56 GPa (c), 2.68 GPa (d), 2.75 GPa (e), 2.9 GPa (f), and 3.2 GPa (g), respectively. The *T*-linear specific heat behavior is labeled by the red regions and is most pronounced at the pressure of zero-field first-order PS-AF transition. (h) The pressure-magnetic-field (*P-B*) phase diagram at the low temperature of 0.4 K. The red region is the *T*-linear specific heat phase, which is most pronounced around the pressure of the zero-field first-order PS-AF transition.

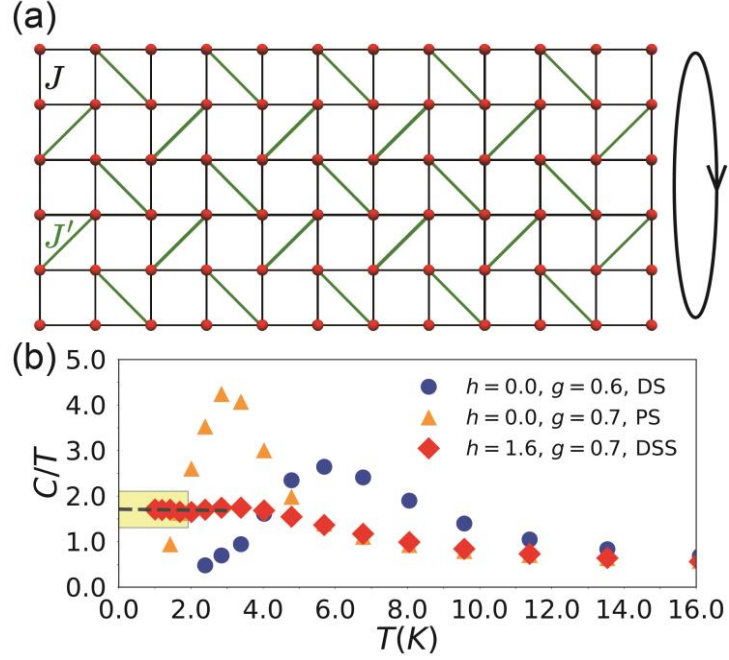

Fig. 4. **XTRG results for the SS lattice.** (a) Schematic of the cylindrical cluster, corresponding to a $3 \times 12 \times 4$ lattice. Black and green bonds represent the inter-dimer coupling $J$ and intra-dimer coupling $J'$ (defined in Eq. 1), respectively. Periodic boundary conditions are applied along the y-direction and open boundary conditions along the $x$-direction. (b) Specific heat divided by temperature, $C/T$, as a function of temperature $T$, computed via XTRG. Results are shown for parameters $g = 0.70$ and $h = 1.6$ (red diamonds, representing the red regime in Fig. 3 (h)), $g = 0.70$ and $h = 0.0$ (orange triangles), and $g = 0.6$ with $h = 1.6$ (blue points). The lowest three temperature points for $g = 0.70$ are used to fit the zero-temperature limit via a linear formula; the resulting extrapolations are plotted as dashed black line. The $T$-linear specific heat region is highlighted by yellow shading.

**Acknowledgements:** This work was supported by the National Key Research and Development Program of China (Grant No. 2021YFA1401800, 2022YFA1403900, No. 2022YFA1403400, 2025YFA1411500 and No. 2021YFA1400400), the NSF of China (Grant Numbers Grants No. U2032214, 12474054, 12122414, 12104487 and 12004419), the Strategic Priority Research Program (B) of the Chinese Academy of Sciences (Grant No. XDB25000000 and No. XDB33000000), the CAS Superconducting Research Project under Grant No.[SCZX-0101], the Synergetic

Extreme Condition User Facility (SECUF https://cstr.cn/31123.02.SECUF), and the K. C. Wong Education Foundation (GJTD-2020-01). J. G. and S. C. are grateful for supports from the Youth Innovation Promotion Association of the CAS (2019008) and the China Postdoctoral Science Foundation (E0BK111). C. H., T. T. W., C. Z. and Z. Y. M. acknowledge the support from the Research Grants Council (RGC) of Hong Kong(Projects Nos. AoE/P-701/20, 17309822 and C7037-22GF, 17302223, 17301924), the ANR/RGC Joint Research Scheme sponsored by the RGC of Hong Kong and French National Research Agency (Project No. A_HKU703/22) and the HKU Seed Funding for Strategic Interdisciplinary Research "Many-body paradigm in quantum moiré material research". We thank HPC2021 system under the Information Technology Services, University of Hong Kong, as well as the Beijing PARATERA Tech CO., Ltd. (URL:https://cloud.paratera.com) for providing HPC resources that have contributed to the research results reported within this paper.